# Spatial distributions of plasma potential and density in electron cyclotron resonance ion source

V. Mironov*, S. Bogomolov, A. Bondarchenko, A. Efremov, V. Loginov, D. Pugachev

*Joint Institute for Nuclear Research, Flerov Laboratory of Nuclear Reactions,
Dubna, Moscow Reg. 141980, Russia*
E-mail: vemironov@jinr.ru

Abstract

The Numerical Advanced Model of Electron Cyclotron Resonance Ion Source (NAM-ECRIS) is applied for studies of the physical processes in the source. Solutions of separately operating electron and ion modules of NAM-ECRIS are matched in iterative way such as to obtain the spatial distributions of the plasma density and of the plasma potential. Results reveal the complicated profiles with the maximized plasma density close to the ECR surface and on the source axis. The ion-trapping potential dips are calculated to be on the level of ~(0.01-0.05) V being located at the plasma density maxima. The highly charged ions are also localized close to the ECR surface. The biased electrode effect is due to an "electron string" along the source axis formed by reflection of electrons from the biased electrode and the extraction aperture. The string makes profiles of the highly charged ions more peaked on the source axis, thus increasing the extracted ion currents.

## Introduction

Numerical modelling of physical processes in Electron Cyclotron Resonance Ion Source (ECRIS) [1] requires studies of the ion and electron dynamics in a dense hot plasma coupled with an intense microwave radiation. Electrons in ECRIS are confined by the magnetic mirror forces and heated by absorption of the microwaves at the electron cyclotron resonance. Electron energies are high enough (~1-100 keV) to effectively ionize the plasma ions up to their high charge states providing that the ion confinement times are sufficiently long (~ 1 ms). The ions are extracted out of the source and form intense beams used e.g. for injection into accelerator facilities.

Our NAM-ECRIS model is a Particle-in-Cell Monte-Carlo Collision group of codes [2 and references therein]; separate modules simulate electron and ion processes in iterative way by exchanging the relevant information between each other. The electron module NAM-ECRIS($e$) traces electron movement in the magnetic field and electron diffusion in the velocity space caused by the ECR heating. The module uses the ion scattering spatially-resolved factors prepared in the ion module NAM-ECRIS($i$). The ion module follows the ion diffusion and ionization by taking into account the electron component's parameters obtained by NAM-ECRIS($e$). The extraction module NAM-ECRIS($x$) calculates the ion extraction out of the source by using the ion spatial and velocity distributions at the extraction aperture imported from NAM-ECRIS($i$).

So far, the ion module was operated in assumptions that electron density is equal to the local ion charge density and that the ions are retarded when crossing the ECR surface by a potential barrier, which value is defined by requiring that the electron and ion losses out of the plasma are equal. After leaving the potential trap, the ions were supposed to be accelerated by the pre-sheath electric fields toward the source walls and extraction.

As the next step in the model development, we calculate the electron density spatial distribution and incorporate it into the ion module, assuring the charge quasi-neutrality by allowing the ions to move in the internal electric fields obtained by solving the 3D Poisson equation. In this way we obtain main parameters of ECRIS plasma with using only one free parameter, namely the electric field amplitude of the resonating microwaves.

Calculations are done for geometry and operational parameters of DECRIS-PM source [3]. The source chamber dimensions are 7 cm in diameter and 23 cm in length. The magnetic field at the chamber walls is 1.1 T, at the injection side of the source on the source axis – 1.34 T, at the extraction side – 1.1 T, and the minimum magnetic field on the axis is 0.42 T. The extraction aperture has a diameter of 1 cm, the biased electrode with a diameter of 3 cm is installed axially at the injection side together with the gas and microwave injection ports. The microwave frequency is 14.5 GHz. We restrict ourselves to simulations of argon plasmas.

The paper is organized in the following way: first, we describe main features of the electron module. Spatial distribution of electron density, electron energy distribution and electron life time are obtained in the module. In the next section, we describe calculations of the ion dynamics, showing the spatial distributions of the plasma potential and ion densities. Also, the biased electrode effect is discussed there. Conclusions are given in the last section.

## Electron dynamics

The methods that we use to simulate the electron dynamics in ECRIS plasma are described elsewhere [4]. Some modifications are made in the algorithm, especially concerning the electron microwave heating.

Electrons are traced as they move in the magnetic field of the source. The particles are elastically scattered in electron-electron and electron-ion collisions according to the density maps imported from the ion module. Also, positions of secondary electron creation in electron-ion collisions are imported from NAM-ECRIS(*i*) to be used as the launching conditions for electrons in the computational domain.

Electrons are reflected from the walls if their energy along the local magnetic field line is less than 50 eV, which is an estimate of the plasma potential drop in a sheath. Close to the biased electrode at the injection side of the source, electrons are reflected back if their longitudinal energy is less than 500 eV.

At the extraction aperture, electrons are retarded by the source extraction potential of 20 keV.

*Interaction with microwaves*

Whenever an electron crosses the ECR surface, it experiences velocity kicks both perpendicular and along the magnetic field line. This resonant diffusion of electrons in velocity space results in an electron heating and in a flux of electrons into the loss cone followed by electron losses on the walls. The relativistic resonance magnetic field is Doppler-shifted in our calculations, $B_{res} = \frac{1}{s} B_0 \gamma (1 - \beta)$, where $B_0 = 0.518\,T$ for 14.5 GHz microwaves, s=1,2 for the 1$^{st}$ and 2$^{nd}$ harmonics of the resonance, $\gamma$ is the relativistic factor, $\beta = \frac{v_\parallel}{v_\Phi}$, $v_\Phi$ is module of the wave phase velocity along magnetic field line, $v_\parallel$ is longitudinal velocity of electron (negative if the wave and electron are moving in opposite directions). The wave phase velocity is calculated from the dispersion relation for the right-hand whistler waves in cold plasma as

$$\frac{c^2}{v_\Phi^2} = n^2 = 1 - \frac{\omega_p^2}{\omega(\omega - \omega_{ce})} = 1 - \frac{n_e^*}{1 - B^*} \quad (1)$$

where n is the refractive index, $\omega_p$, $\omega$ and $\omega_{ce}$ are electron plasma, microwave and electron cyclotron angular frequencies respectively, $n_e^*$ and $B^*$ are the electron density normalized to the critical electron density for 14.5 GHz microwaves (2.6×10$^{12}$ cm$^{-3}$), and the magnetic field normalized to $B_0$ value (the resonant magnetic field with no shift).

In the following, we designate the $B^*$=1 surface as the ECR zone surface. Relativistic and Doppler effects both shift the resonant magnetic fields from the $B^*$=1 value. For $B^*$<1 and $n_e^*$ greater than (1-$B^*$), the phase velocity is an imaginary number and whistler waves are evanescent and non-resonating. Outside the zone, the phase velocity is real for any electron density and no cut-off effects exist there. For $B^*$=1 surface, the phase velocity formally goes to zero, but the electron thermal effects limit it at some level; in our calculations we do not correct Eq.1 for this limitation.

The kick magnitudes are calculated according to Girard et al. [5] with the relativistic corrections from [6]: velocity along the magnetic field $v_\parallel$ increments as

$$\Delta v_\parallel = \frac{eEv_\perp}{m\gamma} \left( \frac{1}{v_\Phi} - \frac{|v_\parallel|}{c^2} \right) t_{eff} \cos(\varphi) \quad (2)$$

where *E* is the resonating microwave electric field amplitude, $t_{eff}$ is the effective time that electron spends in the resonance [7],

$$t_{eff} = min\left\{ \frac{0.71}{\omega} \left( \frac{2\omega}{\omega' v_\perp} \right)^{\frac{2}{3}}, 1.13 \left( \frac{2}{\omega' \omega v_\parallel} \right)^{1/2} \right\} \quad (3)$$

and $\omega' = B_s^{-1}(dB_s/ds)$ is the normalized magnetic field gradient along the field line at the resonance point. Kicks for the perpendicular velocity are

$$\Delta v_\perp = \frac{eE}{m\gamma}(1 - \beta - \frac{v_\perp^2}{c^2})t_{eff}\cos(\varphi) \quad (4)$$

The kick phase φ is random. The microwave amplitude is a free parameter in our calculations. Its value is determined by the level of the injected microwave power, power losses in the plasma and on the walls, as well as direction of the wave propagation after reflections from the plasma and walls. We omit these details at the present version of the model and use the estimate for the changes in the amplitude caused by local refractive index of the plasma (n) as it is suggested in [8]:

$$E \sim E_0 n^{-1/2} \quad (5),$$

where $E_0$ is the wave electric field amplitude in vacuum.

This estimate comes from the relation between the electromagnetic wave energy flux (I) and the wave electric field amplitude: $I = \frac{1}{2}v_\Phi \varepsilon \varepsilon_0 E^2 = \frac{1}{2}nc\varepsilon_0 E^2$ in assumption of negligible variation in the microwave power density while it propagates in plasma. It follows from Eqs.1 and 5 that when the wave is approaching the ECR surface in direction from the higher magnetic field, it is slowed down and its amplitude is decreasing. The larger is the electron density, the more prominent is the amplitude decrease; for the fixed electron density, waves with the higher frequency are damped weaker, which probably explains the frequency scaling effects in ECRIS.

*Electron energy distribution and life time*

The presented data are obtained for the fixed electric field amplitude of microwaves $E_0$ of 100 V/cm. This value is selected such as to obtain the extracted argon ion charge-state distributions close to the experimentally observed ones. Reactions of the plasma parameters to variations in the amplitude and in the gas flow into the source will be reported later.

The typical calculated electron energy distribution for the electrons that stay in the plasma is shown in Fig.1. The spectrum is obtained for the plasma parameters described in the next section after reaching the conformed solutions of the electron and ion modules.

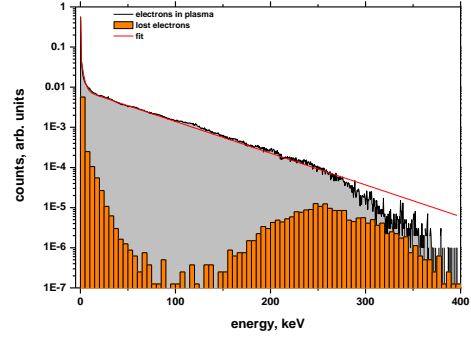

Fig.1 The energy distribution for electrons inside the plasma (grey) and for lost electrons (orange columns).

The distribution is fitted by a sum of three exponentially decaying curves. The high energy part of the distribution can be fitted with two curves with indexes of 2.5 keV (warm component in the energy interval from 100 eV to 10 keV) and 55 keV (hot component with energies above 10 keV), with the warm electrons contributing ~30% into the high-energy part of the spectrum. The low energy part (cold electrons with energies below 100 eV) is fitted by a curve with the decay index of 20 eV. For the presented situation, the cold electron fraction is 30% of the total number of electrons. A knee in the spectrum is seen at energies of ~250 keV (γ≈1.5), for which the relativistically shifted resonant magnetic field start to be comparable to the field close to the extraction.

Energy distribution for the lost electrons (Fig.1, orange columns) differs from the spectrum of electrons in the plasma because of the energy dependence of electron loss rates. Most of the lost electrons have energies below 10 keV, with energetic bump at energies of ~250 keV. Such the bump is seen when measuring the energies of electrons that leave the source through the extraction aperture [9].

Mean energy of the lost electrons is 1 keV. This energy is used for calculation of total power carried by lost electrons, which is close to the microwave power coupled to the source plasma.

The electron life time is defined by two loss mechanisms – diffusion of electrons due to collisions with charged particles in the plasma, and electron losses caused by interaction with the microwaves. The larger is the microwave electric field amplitude, the faster electrons are heated and the collision-induced rates are decreasing (the collision rate depends on the electron velocity as $\sim v_e^{-3}$). At the same time, the microwave-induced rates are increasing with the amplitude. Relative strengths of these channels in our calculations are dependent on the plasma parameters, with the microwave-induced losses dominant for low densities and large electric field amplitudes. For the conditions presented in Fig.1, the electron life time is 0.35 ms, and it is immediately increasing up to 1.9 ms after switching the microwave heating off, indicating the relative importance of the microwave-induced losses.

*Spatial distributions of the electron density*

The electron life time is large compared to the time for electron bouncing along the magnetic field lines and for relatively slow curvature drift across the lines. Most of the cold electrons are trapped by electric fields at the source walls, the biased electrode and the extraction aperture. Energetic electrons are mirror-trapped with the turning points distributed close to their resonance positions. Most of the time particles spend around these turning points, where the velocity along the magnetic field line is small. As the consequence, density of electrons is highest in these regions; for the moderately relativistic plasma in ECRIS, the highest density is around the ECR zone surface. The same issues are discussed in [10].

Trajectories of electrons in the source are illustrated by Fig.2, where slices are shown through trajectories in the transversal (A and B) and in the longitudinal (C) planes. The slice A crosses the point where the magnetic field on axis is minimal (z=12.5 cm). The transversal slice B is calculated at the point where ECR surface crosses the source axis (z=16.5 cm). The longitudinal slice C is oriented perpendicular to the transversal slices such as to cross the hexapole poles (vertically in Figs.2AB). The brighter is a pixel's color in Fig.2, the larger is electron density there.

Losses of the cold electrons are suppressed along the magnetic field lines terminated by the extraction aperture and by the biased electrode. After ECR heating of the trapped electrons, dense compact blobs are formed close to the ECR surface on the axis. The transversal size of the structure is defined by the extraction aperture and by the magnetic field profile; its FWHM is ~8 mm in our case. The longitudinal size depends on the magnetic field gradient around the ECR surface close to the axis, and it is around 1 cm, same for both blobs.

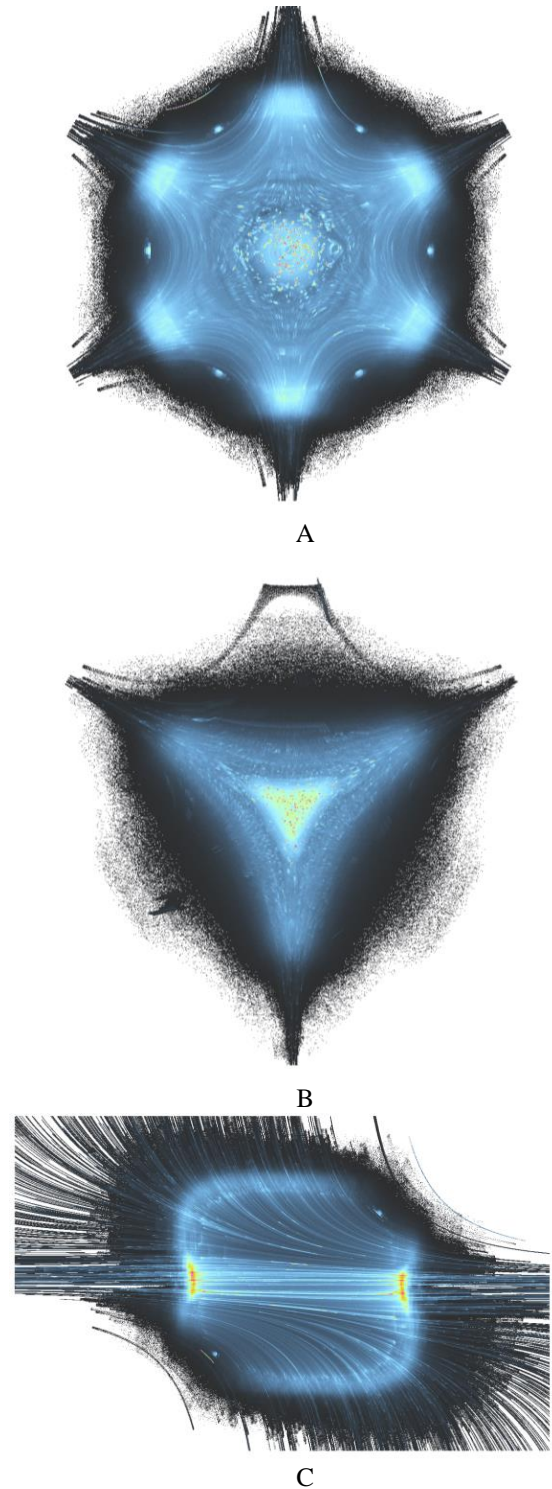

Fig.2 Electron trajectories in the transversal (A and B, z=12.5 and 16.5 cm) and in the longitudinal (C) planes.

We briefly mention here that with using two microwave frequencies for the plasma heating, it is possible to control the dense plasma size along the source axis, which partially explains the experimental observations of the two-frequency heating effects [11].

Transversally, the blobs are triangular stars rotated by 60° in respect to each other at the injection and extraction sides of the ECR zone. When distance from the axis is increasing, the star's arms become more and more inclined toward the source center, being almost parallel to the axis at the center. The blobs are connected by the axial dense bar.

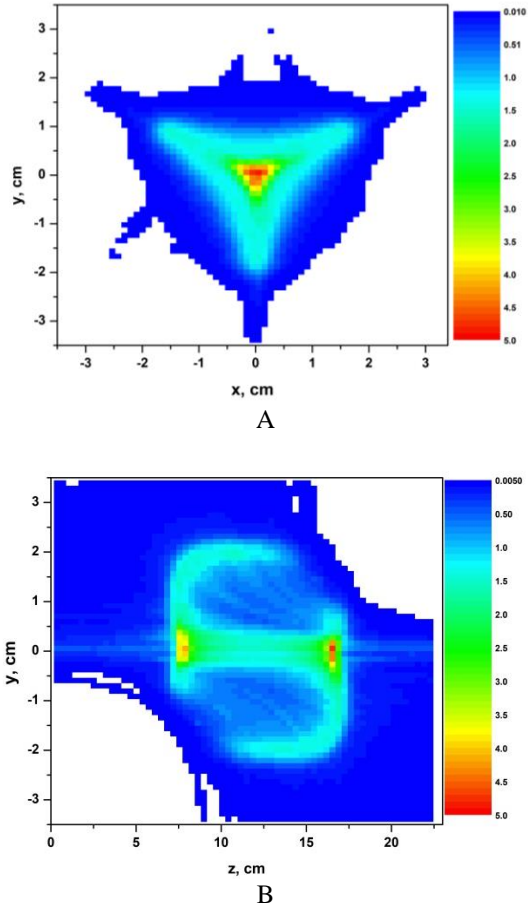

Fig.3 Electron densities in the transversal (A, z=16.5 cm) and longitudinal (B) planes.

From the electron trajectories, we calculate the electron density distributions by using the standard PIC techniques with the under-relaxation algorithm for time-averaging. Typical distributions are shown in Fig.3 in the transversal (z=16.5 cm) and longitudinal planes. The densities are calculated on the computational mesh of 65×65×64 cells in x-, y- and z-directions. The density absolute scale is selected during the iterative procedures that will be described later; for the given dataset, the maximal electron density is $n_{e(max)}=4.5\times10^{12}$ cm$^{-3}$ on the axis and plasma is strongly over-dense in the blobs. The color scale shows the electron density in $10^{12}$ cm$^{-3}$ units.

Perspective drawings of the isosurfaces that limit the plasma with the electron density above $3.0\times10^{11}$ cm$^{-3}$ (A) and $1.1\times10^{11}$ cm$^{-3}$ (B) are shown in Fig.4 for the same plasma parameters as in Fig.3. The plasma boundary is slightly outside the ECR zone with the dumbbell's shape of the densest parts. Narrow axial bars are protruding toward extraction and injection sides of the source from the blobs.

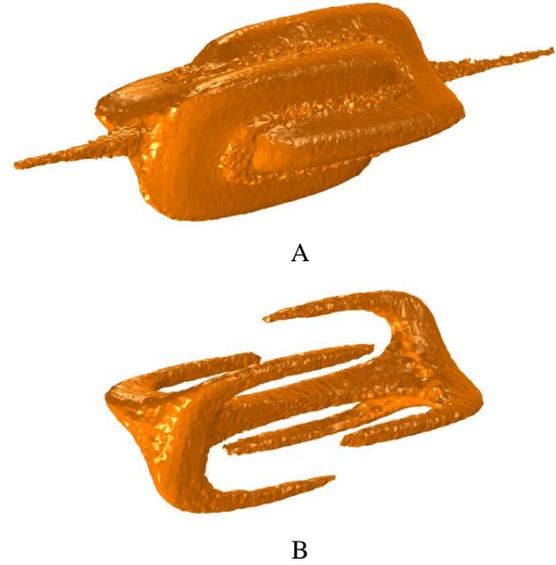

Fig.4 Perspective drawings of the plasma with cuts in electron density from below of $3.0\times10^{11}$ cm$^{-3}$ (A) and $1.1\times10^{12}$ cm$^{-3}$ (B)

The density profiles along the source axis and in transversal direction at z=12 cm (A) and at z=16.5 cm (B) are shown in Fig.5 (longitudinal) and 6 (transversal). The profiles are calculated separately for the "cold" electrons with energies less than 100 eV, for the "warm" electrons with energies from 100 eV to 10 keV and for the "hot" electrons with energies higher than 10 keV. Density of the hot and warm electrons in the blobs is higher by factor of 3 compared to the value at the center of the plasma on axis. The ECR surface coordinates are indicated in Fig.5 and Fig.6A with the dashed orange lines. It is seen that the blobs are located a few millimeters away from the zone positions at the larger magnetic fields.

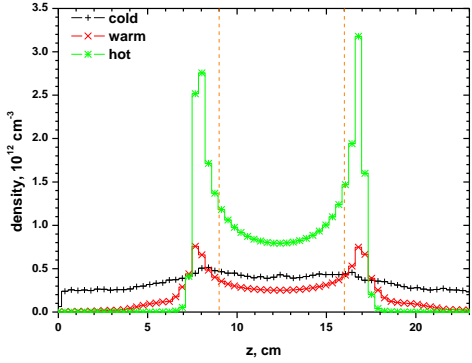

Fig.5 Electron density profiles in axial direction. Black curve represent cold electrons with energies <100 eV, red curve is for the warm electrons with energies from 100 to 10 keV, green curve - for the hot electrons with energies above 10 keV.

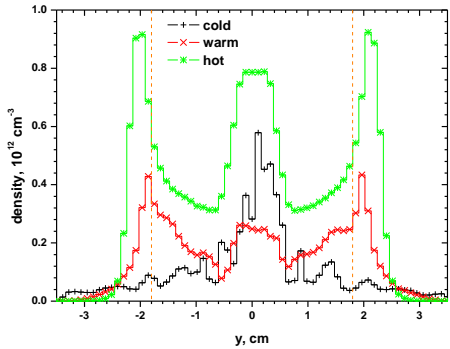

A

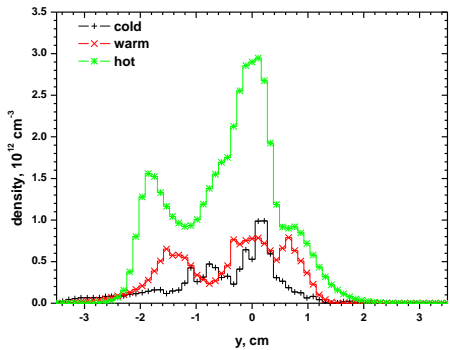

B

Fig.6 Electron density profile in the transversal direction at z=12.5 cm (A) and z=16.5 cm (B) for cold (black), warm (red) and hot (green) electron components.

The profiles are dependent on the electron energy: hot and warm electrons are preferentially localized at the outer parts of plasma and on the axis, while cold electrons are mostly localized in the axial bar, showing the effect of the electrostatic pugging of electrons by the extraction and the biased electrode voltages.

The calculated electron density maps are imported into the ion module of NAM-ECRIS for investigations of the ion dynamics.

## Ion dynamics

*General description and matching of the module solutions*

The ion module uses the PIC-MCC methods to trace movement of ions in the magnetic and internal electric fields of the source [2]. The ions undergo elastic and un-elastic collisions with other ions, neutrals and electrons. Ion neutralization processes on the walls are taken into account either with using the energy accommodation coefficients from [12] or assuming the total energy absorption for the elements heavier or comparable in atomic mass to the wall material (as for Ar+Fe in our case). The electron densities from NAM-ECRIS($e$) are fixed during the specific run and define rates of ionization and heating. We use the electron energy distributions from the electron module to calculate the ionization rates with using the Lotz's cross-sections. The Langevin's rates are used to simulate the charge-exchange collisions.

Each time step, the ion charge densities are calculated on the mesh, as well as the difference between the total ion space charge density and the electron density in a cell. After normalization to the maximal value, the differences are used by 3D Poisson solver to calculate the electric potential $\varphi$. The calculated potential is multiplied by some fixed factor. The larger is this factor, the smaller difference between the ion and electron densities is observed; the factor is selected such as to ensure that the difference is smaller than ~10%. The mesh size is too large to resolve the plasma sheath region, and the potential is obtained without the plasma sheath contribution of ~ +(25-50) V.

The procedure to match solutions of the electron and ion modules is the following: "seed" electron density distribution is imported into the ion module. The ion dynamics is calculated for different statistical weights of computational particles. When reaching stationary conditions in NAM-ECRIS($i$), the extracted ion currents, the gas flow into the source, the total ion flux onto the walls and the ion life time are calculated. The statistical weight is then selected such as to obtain close values of the electron and ion life times. The ion scattering factors for this statistical weight are imported into the electron module, and the cycle is repeated until the solution converges with no noticeable variations in the plasma density profiles and other parameters.

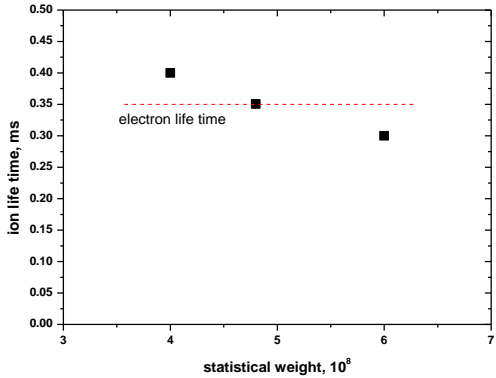

Fig.7 The ion life time as a function of the ion statistical weight.

The typical technical plot of matching the solutions of (*e*) and (*i*) modules is shown in Fig.7. Here, the ion life time is plotted for different statistical weights of ions for the fixed electron density distribution. Increase in the weight (which is an equivalent to increase of gas flow into the source) results in decrease of the ion life time, whereas the electron life time varies insignificantly. The weight of $4.8 \times 10^8$ gives equal electron and ion loss times with the argon gas flow of 0.75 particle-mA.

*Electrostatic potential distributions*

The calculated electric potential distributions along the source axis and in the transversal direction for z=12.5 (red) and 16.5 cm (black) are shown in Fig.8. The potential double layer is formed adjacent to the dense parts of the plasma limited by the ECR surface. The potential minima are formed in the blobs to confine the ions; the values are around 0.05 V along the axis and (0.01-0.02) V in the transversal direction for the given plasma conditions.

The two-dimension plots of the electric potential are shown in Fig.9 in the transversal and longitudinal planes. The transversal distribution follows the plasma symmetry caused by the hexapole component of magnetic field, showing the triangular positive structure with a small ion confining dip on the axis. The longitudinal distribution shows two dips close to the ECR surface at the dense plasma regions. Inside the ECR volume close to the B-minimum position, the electric potential distribution is flat with small depressions close to the peripheral plasma blobs.

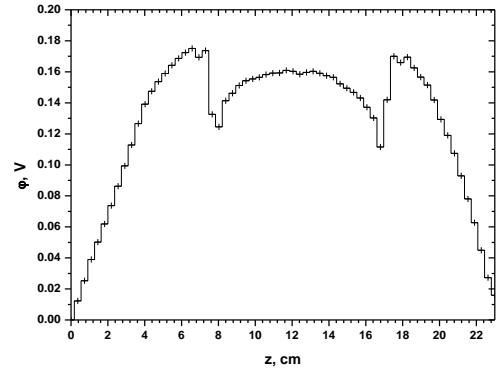

A

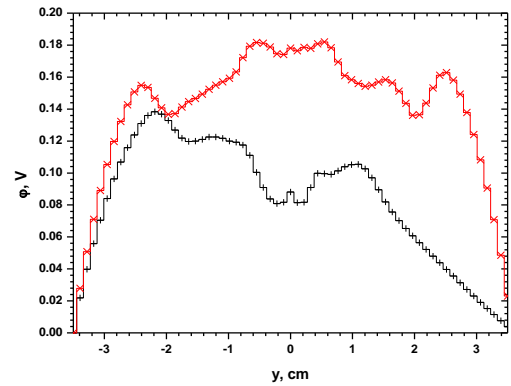

B

Fig.8 Electric potential profiles along the source axis (A) and in the transversal direction (B) at z=12.5 cm (red) and 16.5 cm (black).

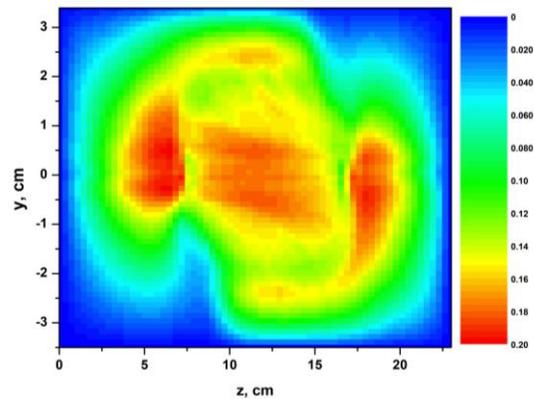

A

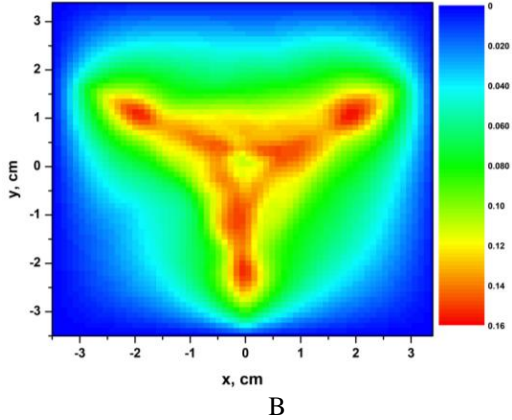

Fig.9 Electric potential in the longitudinal (A) and transversal (B) planes.

We see that the transversal potential dip in the blobs is small and the potential distribution is asymmetric. Potentially, this is the unstable situation: shifts in transversal direction and anomalous radial diffusion of ions can occur in the blobs and in the plasma close to extraction. Experimentally, displacements of the plasma were detected in coincidence with drops in the extracted ion currents [13].

*Spatial distributions of the ion densities*

The potential distributions govern the highly charged ion (HCI) dynamics. These ions are preferentially produced and trapped in minima of the potential that coincide with the maximal electron density regions. The lowly charged ions are distributed mostly outside the ECR zone, with effect of their "burn-out" in the dense regions of the plasma by ionization into the higher charge states. Trajectories of $Ar^{8+}$ and $Ar^{1+}$ ions are shown in Fig.10 and Fig.11 respectively. The same color scheme is used as for Fig.2, with the brighter colors indicating that more ions crossed a pixel.

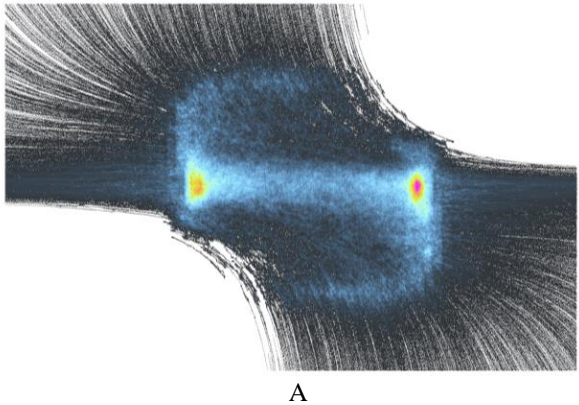

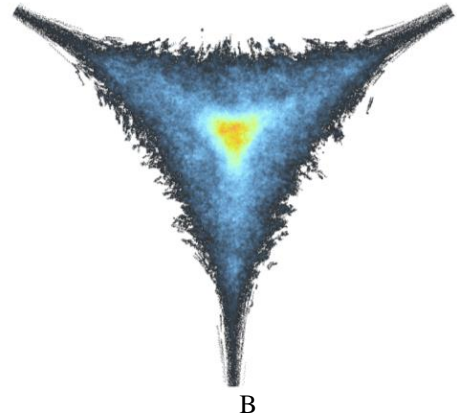

Fig.10 Trajectories of $Ar^{8+}$ ions in longitudinal (A) and transversal (B, z=16.5 cm) planes.

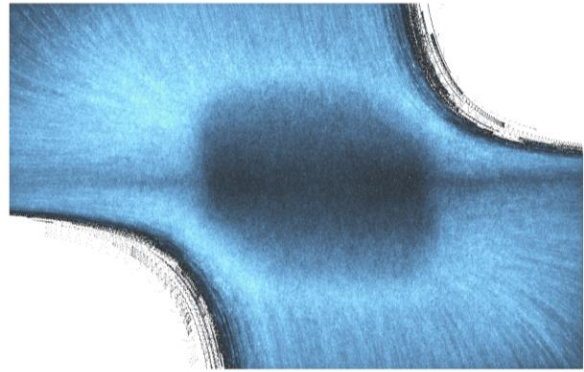

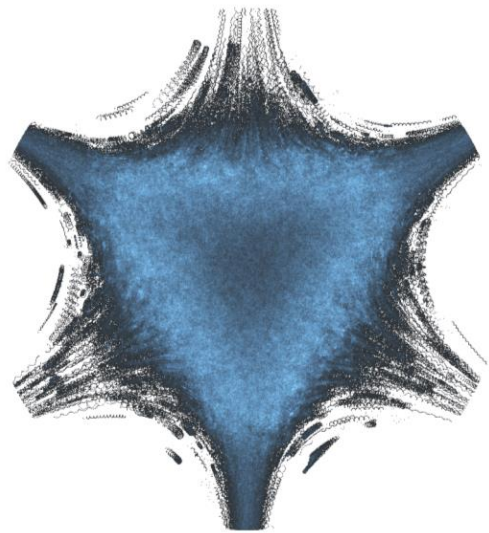

Fig.11 Trajectories of $Ar^{1+}$ ions in longitudinal (A) and transversal (B, z=16.5 cm) planes.

Ion temperatures weakly depend on the ion charge state because of a strong energy equilibration in the ion-ion collisions. The temperature for $Ar^{8+}$ ions is calculated as 0.15 eV, while the coldest $Ar^{1+}$ ions have the temperature of 0.08 eV, close to the longitudinal dip value.

From the total flux of ions toward the source walls and into extraction, we estimate the electron flux and, by using the mean energy of the lost electrons of 1 keV from the electron module, calculate that the power flux out of the plasma is ~50 W. This value should be compared with ~500 W of the total power of microwaves injected into the source.

Strong localization of HCI on the source axis and in the plasma blobs influences the profiles of extracted ions. Positions of those ions that are lost at the extraction electrode are shown in Fig.12. Three-arm pattern is always observed in experiments and had been reproduced already in our previous publications [2]. The main feature in the Fig.12 is a compact spot in the center of distribution and relatively intense strips along the star arms.

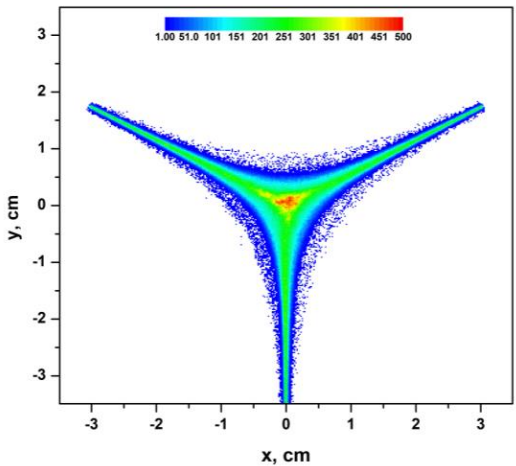

Fig.12 Positions of the lost ions on the extraction electrode.

The higher is the ion charge state, the smaller is the size of ion distribution at the extraction aperture. For the $Ar^{10+}$ ions, the mean radius of particles that pass through the extraction aperture is 2.6 mm, 25% smaller than 3.5 mm for the uniformly distributed ions. The result is that the magnetic emittance term for these ions is reduced to 54% of the value for the uniform distribution. Such deviations of the measured emittances of the highly charged ions from the estimated values were reported elsewhere [14].

*The biased electrode effect*

We demonstrated that electrostatic trapping of electrons along the magnetic field lines by the biased electrode and the extraction voltage boosts the plasma density along the source axis. To observe the effect more clearly, electron density maps are obtained with setting no electron-retarding voltage on the biased electrode such that electrons are reflected there by the sheath potential only. The extraction voltage is not changed such that the cold and warm electrons are still reflected at the extraction aperture. The densities are obtained in the same iterative way as it is described in the previous section.

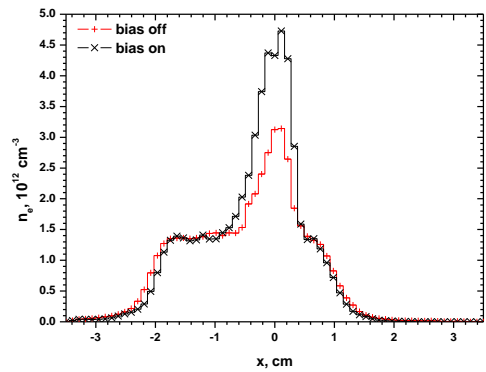

Fig.17 Electron densities in transversal direction at z=16.5 cm with the biased electrode on (black) and off (red).

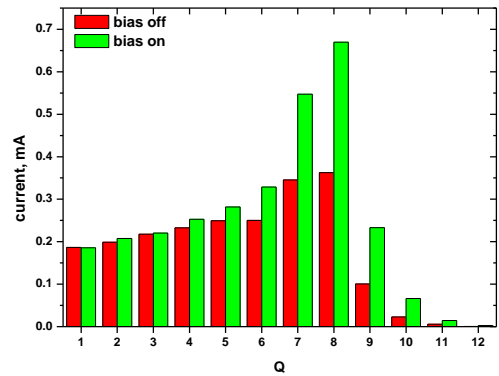

Fig.18 Charge state distribution of extracted argon ions with (green) and without (red) the biased electrode voltage switched on.

Confinement of electrons along the source axis is weakened in these conditions and the electron density in the blobs is decreased. Profile of total electron density in the transversal direction for z=16.5 cm is shown in Figs.17 as the red curve. The gas flow is of

0.75 particle-mA, i.e. the plasma input parameters are the same as for the data shown previously. The distribution is compared to the data with the biased voltage switched on (black curve, which is obtained by summation of densities for all electron components from Fig.6B). It is seen that the electron density in the blob on axis is decreased by ~30% with the biased electrode voltage off, mostly because of the reduced content of the cold and warm electrons. At the same time, electron density at the outer parts of the plasma remains unchanged.

The extracted argon ion charge-state distributions with and without biasing the electrode are compared in Fig.18. Currents of the highly charged ions above 6+ charge state drop noticeably, while currents of the moderately and lowly charged ions are not affected. Changes in the source output are consistent with the experimental observations [15,16].

The "electron string" formation and increase of the plasma density on axis requires trapping potentials both at the injection and extraction sides of the source: we expect different plasma parameters with and without putting the extraction voltage on the source.

## Conclusions

Mirror-trapped electrons with energies of ~(10-50) keV spend most of time at their turning points close to the ECR surface. The result is that the plasma density is maximized there. Electrostatic trapping of electrons in between the biased electrode and extraction aperture boosts the plasma density on the source axis, with formation of two dense blobs at intersection of the "electron string" with the ECR surface. Plasma density in these blobs can be larger by factor of 2-3 of the density in the source center around the minimum of the magnetic field. Highly charged ions are electrostatically trapped in these regions. Small transversal size of the blobs means a strong axial localization of the highly charged ions at the extraction aperture and reduced emittances of the extracted ion beams.

The concept of ECRIS plasma as a combination of two compact over-dense blobs surrounded by a hot shell located around the ECR surface can explain experimental observations of source responses to changes in the magnetic field gradients at the ECR surface and the two-frequency heating effects.

Electron dynamics calculations in our model suffer from assumptions of uniform electric field amplitude of the resonating microwaves ($E_0$) outside the plasma and of the amplitude's reduction in correspondence to the local refractive index of plasma. In real conditions the microwaves are absorbed, reflected and refracted by the plasma such that we expect $E_0$ value varying inside the source chamber. More sophisticated modelling of the electron heating process can result in different electron energy distributions and electron life times and thus in different parameters of the ion component. Nonetheless, the presented version of the NAM-ECRIS code touches important aspects of ECRIS operation. Adjustment of only one free parameter is needed to reproduce the experimentally measured extracted ion currents. The biased electrode effect is explained in the model as the consequence of increased plasma density in the blobs.

[1] R.Geller, "Electron Cyclotron Resonance Ion Sources and ECR Plasma", (Bristol: Institute of Physics), 1996
[2] V. Mironov, S. Bogomolov, A. Bondarchenko, A. Efremov, K. Kuzmenkov and V. Loginov, "Simulations of ECRIS performance for different working materials", JINST, **13**, C12002 (2018); doi:10.1088/1748-0221/13/12/C12002
[3] G.G. Gulbekian, et al., "Proposed design of axial injection system for the DC-280 cyclotron", Phys. Part. Nuclei Lett., vol. 11, p.763, 2014, doi:10.1134/S1547477114060028
[4] V. Mironov, S. Bogomolov, A. Bondarchenko, A. Efremov and V. Loginov, "Some aspects of electron dynamics in electron cyclotron resonance ion sources", Plasma Sources Science and Technology, **26**, 075002 (2017), doi:10.1088/1361-6595/aa7296
[5] A. Girard, C. Perret, G. Melin, and C. Lécot, "Modeling of electron-cyclotron-resonance ion source and scaling laws", Review of Scientific Instruments, **69**, 1100 (1998); doi:10.1063/1.1148588
[6] K. R. Chu, "The electron cyclotron maser", Rev. Mod. Phys., **76**, 489 (2004); doi:10.1103/RevModPhys.76.489
[7] M.A. Lieberman and A.J. Lichtenberg, "Theory of electron cyclotron resonance heating. II. Long time and stochastic effects", Plasma Phys., **15,** 125 (1973); doi:10.1088/0032-1028/15/2/006
[8] Gareth Guest, "Electron Cyclotron Heating of Plasmas", (Weinheim, Germany: Wiley-VCH-Verlag-GmbH), 2009; doi:10.1002/9783527627882
[9] I. Izotov, O. Tarvainen, V. Skalyga, D. Mansfeld, T. Kalvas, H. Koivisto and R. Kronholm, "Measurement of the energy distribution of electrons escaping minimum-B ECR plasmas", Plasma Sources Sci. Technol., **27,** 025012 (2018);doi: 10.1088/1361-6595/aaac14

[10] D. Mascali, L. Celona, S. Gammino, G. Castro, R. Miracoli, F.P. Romano, L. Malferrari, F. Odorici, R.Rizzoli, G.P. Veronese and T. Serafino, "An investigation on the formation of suprathermal electrons in a B-min ECR machine and a novel method for their damping", Plasma Sources Sci. Technol., **22**, 065006 (2013); doi:10.1088/0963-0252/22/6/065006

[11] Z. Q. Xie and C. M. Lyneis, "Two-frequency plasma heating in a high charge state electron cyclotron resonance ion source", Review of Scientific Instruments, **66**, 4218 (1995); doi:10.1063/1.1145372

[12] Frank O. Goodman, "Thermal accommodation coefficients", J. Phys. Chem., **84**, 1431 (1980); doi: 10.1021/j100449a002

[13] V. Mironov, J.P.M. Beijers, S. Brandenburg, H.R. Kremers, J. Mulder, S. Saminathan, "ECR Ion Source Development at the AGOR Facility", In: Cyclotrons and their applications. Proceedings, 19th International Conference, Cyclotrons 2010, Lanzhou, China, September 6-10, 2010; http://accelconf.web.cern.ch/AccelConf/Cyclotrons2010/papers/mopcp053.pdf

[14] D. Leitner, D. Winklehner and M. Strohmeier, "Ion beam properties for ECR ion source injector systems", JINST, **6**, P07010 (2011); doi:10.1088/1748-0221/6/07/P07010

[15] S. Gammino, J. Sijbring, and A. G. Drentje, "Experiment with a biased disk at the K.V.I. ECRIS", Review of Scientific Instruments, **63**, 2872 (1992); doi:10.1063/1.1142782

[16] S. Runkel, O. Hohn, K. E. Stiebing, A. Schempp, H. Schmidt-Böcking, V. Mironov and G. Shirkov, "Time resolved experiments at the Frankfurt 14 GHz electron cyclotron resonance ion source", Review of Scientific Instruments, **71**, 912 (2000); doi:10.1063/1.1150343